# The wave equation

Benjamin Titov, student of the MSU physical faculty.
(mks2002@bk.ru)

### The introduction.

As well known, the wave equation is a postulate, the inviolability which is undoubtable. The most obvious confirmation of this equation for Schrödinger and for his contemporaries became the fact, that it describes the hydrogen energy levels with good accuracy. In this article we attempt to find reasons, why it was possible for Schrödinger to "guess" it so precisely. We shall formulate the postulates, from which this equation can be obtained. In the end it will be understandable, that it is the most natural law describing the motion of a system. The postulates themselves are simple and apparent: the first and the fifth postulate set the definition of the coordinate and the definition of its evolution. The second and the fourth postulate define the momentum and define its evolution. The third postulate sets the superposition principle. On the basis of these statements it is possible to obtain the wave equation.

### The main part.

In the beginning we shall consider some connections between this equation and the Maxwell equations, which, as we shall see further, are much more deep, than it seems at the first time. The Maxwell equations are those:

$$\begin{cases} rot\,\vec{H} = \frac{1}{c}\frac{\partial \vec{E}}{\partial t} + \frac{4\pi}{c}\vec{j} \\ rot\,\vec{E} = -\frac{1}{c}\frac{\partial \vec{H}}{\partial t} \\ div\,\vec{E} = 4\pi\rho \\ div\vec{H} = 0 \end{cases}$$

(1)

The very relevant feature is the fact, that, knowing the configuration $(\vec{E}, \vec{H})$ in the initial moment of time, we know it and at all subsequent moments. From some point of view it "contradicts" the intuition: it is not enough for us to know initial coordinates of a system, it is necessary to know its momentum also in order to know, how the motion goes. From the intuitive point of view is strange that the "photograph" of a system at the initial point does not give us its full description. As we shall see further, the quantum theory applied in the case of mechanical movement, will give a positive answer on this problem, having many similar features with electrodynamics.

In the Maxwell theory the energy density is as follows:

$$\frac{dW}{dV} = \frac{\left|\vec{E}\right|^2 + \left|\vec{H}\right|^2}{8\pi}$$

(2)

And the whole energy is an integral:

$$W = \int \frac{|\vec{E}|^2 + |\vec{H}|^2}{8\pi} dV = \frac{1}{8\pi} \int |\vec{E}|^2 dV + \frac{1}{8\pi} \int |\vec{H}|^2 dV \tag{3}$$

The energy can be calculated in another way. We can decompose $\vec{E}$ and $\vec{H}$ into harmonics and integrate them in the wave vector space. The simple physical sense of this procedure is that each harmonic keeps in itself a definite energy, and the integration counts the total energy. From the mathematical point of view it is enough to use of the Perceval equation:

$$W = \frac{1}{8\pi} \int |\vec{E}(\vec{k})|^2 dV + \frac{1}{8\pi} \int |\vec{H}(\vec{k})|^2 dV, \tag{4}$$

where $\vec{E}(\vec{k}) = \frac{1}{(2\pi)^{3/2}} \int \vec{E}(\vec{r}) e^{-i\vec{k}\vec{r}} dV$

$$\vec{H}(\vec{k}) = \frac{1}{(2\pi)^{3/2}} \int \vec{H}(\vec{r}) e^{-i\vec{k}\vec{r}} dV$$

Already in Maxwell theory it is possible is to define the space of functions
$$\Psi : \vec{k} \to (\vec{E}(\vec{k}), \vec{H}(\vec{k}))$$
And then the energy will be calculated as

$$E = (\Psi, f\Psi) \tag{5}$$

Where
$$f(\vec{E}(\vec{k}), \vec{H}(\vec{k})) = \frac{1}{8\pi} (\vec{E}(\vec{k}), \vec{H}(\vec{k}))$$

One fact here is relevant: each harmonic bears in itself an energy, which equals
$$\frac{|\vec{E}(\vec{k})|^2 + |\vec{H}(\vec{k})|^2}{8\pi} \tag{6}$$

In a quantum case we shall receive similar expression.

And now we shall formulate directly the postulates, from which the will make the derivation. Before writing of the first postulate we shall imagine an experiment. The words, which will be said are not postulates, they will be a little bit later. They must be taken as some arbitrary discussion. Let any particles (for example, the electrons) fly through some hole and fall on a screen, and there we see the diffraction picture.

This experiment leads us to the wave hypothesis, which says, that our particles are described by some vector function $\vec{\Phi}$, similar to the electric field amplitude, which exhibits wave properties. We consider this function to be complex valued (its real part describes the wave itself).

The main result of the given experiment is that our wave, passing through the hole can be described accurately enough with the Huygens theory, which says that the wave is a superposition of point sources arranged on a surface of its constant phase. Let's remark, that these reasoning do not claim at all for strictness. The postulates will be strict. The last thing we must note is, that the theory of Huygens gives one more important result. The properties of the diffraction picture are very similar to those, which we obtain in the light diffraction experiments. Actually it is possible to suspect, that the system (which is described by a vector function $\vec{\Phi}$), and the value, which describes the diffraction picture is $|\vec{\Phi}|^2$ (analogous the electromagnetic field amplitude $(\vec{E}, \vec{H})$, and to the function $|\vec{E}|^2 + |\vec{H}|^2$, characterizing the diffraction picture).

Therefore our first postulate will be about it, and, returning to the experiment, we see, that the distribution of intensity is possible to understand as the flux density of come particles. But the previous characteristic we have connected with $|\vec{\Phi}|^2$. In addition, the flux density is proportional to the spatial density. Therefore it is possible to identify $|\vec{\Phi}|^2$ with spatial density.

We come to the statement, which we will set as an axiom:

THE POSTULATE 1 Our system is described completely with a complex valued vector function $\vec{\Phi}(\vec{r}, t)$.

Probability density to find a system in the given point is $|\vec{\Phi}(\vec{r}, t)|^2$.

From this postulate the normalization condition $\int |\vec{\Phi}(\vec{r}, t)|^2 dV = 1$ immediately follows.

The second corollary of a postulate is that if we want to calculate the mean coordinate $\vec{r}$ of our system, it is enough to write the product of the vector $\vec{r}$ and the distribution function $|\vec{\Phi}|^2$ and take a volume integral:

$$<\vec{r}> = \int \vec{r} |\vec{\Phi}(\vec{r}, t)|^2 dV$$

Let's rewrite this equation in another way:

$$<\vec{r}> = \int \Phi_\alpha^*(\vec{r}, t) \, \vec{r} \, \Phi_\alpha(\vec{r}, t) \, dV$$

That is

$$<\vec{r}> = (\Phi_\alpha, \vec{r} \Phi_\alpha) \quad (7)$$

Therefore to calculate the coordinate $<\vec{r}>$ of a system, we must average the operator

$$\hat{\vec{r}} = \begin{pmatrix} \hat{x} \\ \hat{y} \\ \hat{z} \end{pmatrix} \text{ over } \vec{\Phi}(\vec{r}, t).$$

Now the coordinate operator is put into operation. Now let we want to calculate a mean potential energy of a system. Let the system be in a potential $U(\vec{r})$. We act similarly. It is necessary to average a function $U(\vec{r})$ over the distribution function $|\vec{\Phi}(\vec{r}, t)|^2$. And we shall receive:

$$<U> = \int U(\vec{r}) |\vec{\Phi}(\vec{r}, t)|^2 dV$$

Or, the same,

$$<U> = \int \Phi_\alpha^*(\vec{r}, t) U(\vec{r}) \Phi_\alpha(\vec{r}, t) \, dV$$

That is,

$$<U> = (\Phi_\alpha, U \Phi_\alpha) \quad (8)$$

Now we have token into account the potential energy operator – the operator of multiplying on $U(\vec{r})$.

Now it's natural to evaluate the average speed. We can take $v(\vec{r})$ and then average it over $|\vec{\Phi}(\vec{r},t)|^2$. But the problem is that WE HAVE NO WAY TO FIND OUT $v(\vec{r})$. Here we have the impossibility to calculate the speed characteristics in such a way. Therefore it is necessary to enter into consideration new experiments, and on their basis to write new postulates.

Now we let us consider an electron diffraction experiment from the different point of view. Let electrons radiated by some source, fall on a crystal and diffract. And now we allow, that we want to measure the pressure which they give.

This experiment gives two very important results. At first, we see, that the pressure is proportional to the flux density of electrons (flux is a characteristic, which one is easy to measure. It, roughly speaking, is proportional to a current through a filament which emits particles). It is possible to analyze this fact from the theoretical point of view: the flux of momentum is proportional to the particle flux, the constant of proportionality is the momentum of a particle.

It is easy to see, that the particle flux is proportional to spatial density. This is the value, which $|\vec{\Phi}|^2$ also characterizes. So we see, that **in our experiment** the momentum is proportional to $|\vec{\Phi}|^2$:

$$\vec{p} \sim |\vec{\Phi}|^2$$

Now let us remember the de Broglie hypothesis. It's statement is that we can link a wave vector $\vec{k}$ to a particle moving with a some speed In our experiment all electrons move with the same speed, and they are characterized by a vector function $\vec{\Phi}$, therefore, it's very natural to suppose function $\vec{\Phi}$ has approximately one harmonic in it's spectrum: $\vec{\Phi} \sim e^{i\vec{k}\vec{r}} \vec{\Phi}(\vec{k})$ within the limits of area of our experiment, $\vec{\Phi}(\vec{k})$ is an amplitude of the corresponding harmonic.

Comparing last two equalities, we see, that

$$\vec{p} \sim |\vec{\Phi}(\vec{k})|^2$$

The second conclusion, which one we can make from our experiment also is, that the pressure provided by electrons, is proportional to the module of a wave vector $|\vec{k}|$. It is easy to see that the pressure is directed the same as $\vec{k}$. However, the empirical formula $\vec{p} = \hbar \vec{k}$ tells us about this proportionality:

$$\vec{p} \sim \vec{k}$$

Therefore, connecting the last two equalities, we see, that

$$\vec{p} \sim \vec{k} |\Phi(\vec{k})|^2$$

In order to find the total pressure we must take an integral:

$$\vec{p} \sim \int \vec{k} |\Phi(\vec{k})|^2 d\vec{k}$$

Or,

$$\vec{p} \sim \int \Phi_\alpha^*(\vec{k}) \vec{k} \, \Phi_\alpha(\vec{k}) \, d\vec{k}$$

Let's use again the Parseval equation:

$$\vec{p} \sim \int \Phi_\alpha^*(\vec{r}) \, (-i \frac{\partial}{\partial \vec{r}}) \, \Phi_\alpha(\vec{r}) \, d\vec{r}$$

The proportionality constant is named as Planck's constant $\hbar$.

So, we have:
$$\vec{p} = \hbar \int \Phi_\alpha^*(\vec{r})\,(-i\frac{\partial}{\partial \vec{r}})\,\Phi_\alpha(\vec{r})\,d\vec{r}$$

Or, the same,
$$<\vec{p}> = (\Phi_\alpha, -i\hbar \frac{\partial}{\partial \vec{r}} \Phi_\alpha)$$

The operator of a momentum $\hat{\vec{p}} = -i\hbar \frac{\partial}{\partial \vec{r}}$ is taken into consideration. And now after those speculations let us formulate the postulate. In fact, it is based on the de Broglie hypothesis.

THE POSTULATE 2. The momentum of a system is an integral:
$$\vec{p} = \int \hbar \vec{k}\,|\Phi(\vec{k})|^2\,d\vec{k} \qquad (9)$$

As we have seen now, this postulate gives the following:
$$\vec{p} = \int \Phi_\alpha^*(\vec{k})\,\hbar\vec{k}\,\Phi_\alpha(\vec{k})\,d\vec{k}$$

That is,
$$\vec{p} = \int \Phi_\alpha^*(\vec{r})\,(-i\hbar\frac{\partial}{\partial \vec{r}})\,\Phi_\alpha(\vec{r})\,d\vec{r}$$

So,
$$<\vec{p}> = (\Phi_\alpha, -i\hbar \frac{\partial}{\partial \vec{r}} \Phi_\alpha) \qquad (10)$$

And now we should determine the most important thing - the dynamics of a system. In our terms (see Postulate 1) it is necessary to understand the evolution law of vector function $\vec{\Phi}(\vec{r},t)$.

Let's remark, that by introduction of the operator of momentum, operator of coordinate, etc., we have not advanced in this problem at all. We have understood how to calculate physical values, but we do not know how they change. And also we haven't considered a very important operator, which will connect kinematics and dynamics - the operator of force. To do it we will act the same way.

We know that the spatial density is $|\vec{\Phi}(\vec{r},t)|^2$. We also know that the force acting on a particle is $-\frac{\partial U}{\partial \vec{r}}$. Therefore in order to calculate the force we must write:
$$<\vec{F}> = \int -\frac{\partial U}{\partial \vec{r}}\,|\vec{\Phi}(\vec{r},t)|^2\,dV$$

Or,
$$<\vec{F}> = \int \Phi_\alpha^*(\vec{r},t)\,(-\frac{\partial U}{\partial \vec{r}})\,\Phi_\alpha(\vec{r},t)\,dV$$

Then
$$<\vec{F}> = (\Phi_\alpha, -\frac{\partial U}{\partial \vec{r}} \Phi_\alpha) \qquad (11)$$

Thereby we have entered the operator of force $\hat{\vec{F}} = -\dfrac{\partial U}{\partial \vec{r}}$. After that let us see directly, how $\vec{\Phi}(\vec{r},t)$ changes. As known, the system is completely set by this function. This implies that the system position at the moment $t > 0$ depends ONLY on a system position in the previous moment:

$$\vec{\Phi}(\vec{r}, t+dt) = L\vec{\Phi}(\vec{r},t)$$

We know nothing about the operator $L$ (generally, we do not know, whether the operator $L$ is linear and the more so self conjugated). Let's enter the new operator $A$, determined by the equation $L = 1 + A\,dt$. And then $\vec{\Phi}(\vec{r}, t+dt) = (1 + A\,dt)\vec{\Phi}(\vec{r},t)$. And we come to the fact telling us that the change of a system is determined by an equation:

$$\frac{\partial \vec{\Phi}}{\partial t} = A\vec{\Phi} \qquad (12)$$

See also Appendix 2, where this equation is proved more strictly. The situation is the same as in a classic case. There the system is completely determined by a coordinate and momentum, the change of a system is determined by the first order equations $\dfrac{dq}{dt} = \dfrac{\partial H}{\partial p}, \dfrac{dp}{dt} = -\dfrac{\partial H}{\partial q}$. Let us return to our case. As it was mentioned, we don't know even the linearity o four operators. Strictly speaking, the linearity includes two properties:

$$A(\Phi_1 + \Phi_2) = A\Phi_1 + A\Phi_2$$
$$A(\alpha\,\Phi) = \alpha\,A(\Phi)$$

The first property immediately follows from a principle of superposition (which will be set as a postulate). The second property can be reasoned in the following way. Initially we have normalization condition $\|\vec{\Phi}\| = 1$ (see the corollary of Postulate 1). Therefore if we multiply it by a constant, this condition will be broken: $\|\alpha\Phi\| = |\alpha|$.

Thus, strictly speaking, we have no the right to multiply by a constant, and the operator $A$ can act only on the normalized vectors, and the change of the system should save this condition. Therefore if operator $A$ initially acted only on the normalized vectors, WE CAN DETERMINE IT, so it could act and on unnormalized vectors. We determine by the most natural image:

$$A(\alpha\,\Phi) \equiv \alpha\,A(\Phi)$$

And, thereby, second property of linearity is satisfacted. Therefore, the dynamical equation of a system can be written:

$$\frac{\partial \vec{\Phi}}{\partial t} = \hat{A}\vec{\Phi},$$

where $\hat{A}$ is a **linear operator**

We have used the superposition principle. Let's formulate it as the separate postulate:

THE POSTULATE 3 («principle of "superposition"») If $\vec{\Phi}_1(\vec{r},t)$ and $\vec{\Phi}_2(\vec{r},t)$ - are possible laws of the system evolution, then $\vec{\Phi}_1(\vec{r},t) + \vec{\Phi}_2(\vec{r},t)$ is also the law system evolution (more strictly, we must write not $\vec{\Phi}_1(\vec{r},t) + \vec{\Phi}_2(\vec{r},t)$, but $\dfrac{\vec{\Phi}_1(\vec{r},t) + \vec{\Phi}_2(\vec{r},t)}{\|\vec{\Phi}_1(\vec{r},t) + \vec{\Phi}_2(\vec{r},t)\|}$ in order to satisfy the normalization condition).

Let's consider the last equation and prove that operator $\hat{A}$ is self-conjugated. We do it, proposing that this operator is time-independent (in Appendix 2 we prove it without using this condition, but a little bit longer). In our situation the time-independence means that it does not matter, at which moment of time our system began to evolve: the development will go under the same law. According to our last equation the development is characterized by operator $\hat{A}$. The time independence means that

$$\hat{A}(t+dt) = \hat{A}(t)$$

In other words, $\hat{A}(t)$ is a constant:

$$\hat{A}(t) = const \qquad (13)$$

Using this condition it is possible to solve the evolution equation (12):

$$\vec{\Phi} = e^{\hat{A}t}\vec{\Phi}_0 \qquad (14)$$

But our states are obviously normed: $\|\vec{\Phi}\| = 1$ and $\|\vec{\Phi}_0\| = 1$. Therefore it is easy to see, that the operator $e^{\hat{A}t}$ saves the norm. This means that $e^{\hat{A}t}$ is a unitary operator (according to the definition).

As known, if $e^{\hat{A}t}$ is unitary, then the operator $\hat{A}$ is anti-hermittean (that is $\hat{A}^* = -\hat{A}$). See the proof in appendix 1. Now let us consider the operator $\hat{B} = i\hbar\hat{A}$. This operator will be hermittean (that is, $\hat{B}^* = \hat{B}$). Than equation (12) can be rewritten in a standard way:

$$i\hbar\frac{\partial\vec{\Phi}}{\partial t} = \hat{B}\vec{\Phi}, \qquad (15)$$

where $\hat{B}$ is a **Hermittean (self conjugated) linear operator**.

The last equation is very important. With the help of it we can introduce the notion of derivative. The definition is standard, well-known:

$$\frac{d\hat{X}}{dt} = \frac{\partial\hat{X}}{\partial t} + \frac{i}{\hbar}\left[\hat{B}, \hat{X}\right] \qquad (16)$$

This equation is proved the same way as it is done in the quantum theory. We consider the time derivative of a mean value:

$$\frac{d}{dt}(\vec{\Phi}, \hat{X}\vec{\Phi}) = (\frac{\partial\vec{\Phi}}{\partial t}, \hat{X}\vec{\Phi}) + (\vec{\Phi}, \frac{\partial\hat{X}}{\partial t}\vec{\Phi}) + (\vec{\Phi}, \hat{X}\frac{\partial\vec{\Phi}}{\partial t})$$

Using (15), we obtain:

$$\frac{d}{dt}(\vec{\Phi}, \hat{X}\vec{\Phi}) = (\frac{1}{i\hbar}\hat{B}\vec{\Phi}, \hat{X}\vec{\Phi}) + (\vec{\Phi}, \frac{\partial\hat{X}}{\partial t}\vec{\Phi}) + (\vec{\Phi}, \hat{X}\frac{1}{i\hbar}\hat{B}\vec{\Phi}) = \frac{i}{\hbar}(\hat{B}\vec{\Phi}, \hat{X}\vec{\Phi}) + (\vec{\Phi}, \frac{\partial\hat{X}}{\partial t}\vec{\Phi}) - \frac{i}{\hbar}(\vec{\Phi}, \hat{X}\hat{B}\vec{\Phi})$$

But $\hat{B}$ is self-conjugated. That's why

$$\frac{d}{dt}(\vec{\Phi}, \hat{X}\vec{\Phi}) = \frac{i}{\hbar}(\vec{\Phi}, \hat{B}\hat{X}\vec{\Phi}) + (\vec{\Phi}, \frac{\partial\hat{X}}{\partial t}\vec{\Phi}) - \frac{i}{\hbar}(\vec{\Phi}, \hat{X}\hat{B}\vec{\Phi}) = (\vec{\Phi}, (\frac{i}{\hbar}\hat{B}\hat{X} + \frac{\partial\hat{X}}{\partial t} - \frac{i}{\hbar}\hat{X}\hat{B})\vec{\Phi})$$

Therefore we have:

$$\frac{d}{dt}(\vec{\Phi}, \hat{X}\vec{\Phi}) = (\vec{\Phi}, (\frac{i}{\hbar}\hat{B}\hat{X} + \frac{\partial\hat{X}}{\partial t} - \frac{i}{\hbar}\hat{X}\hat{B})\vec{\Phi}) \qquad (17)$$

Here we see that function $\vec{\Phi}$ is arbitrary. So we can write that the derivative of operator $\hat{X}$ can be evaluated standartly:

$$\frac{d\hat{X}}{dt} = \frac{i}{\hbar}\hat{B}\hat{X} + \frac{\partial \hat{X}}{\partial t} - \frac{i}{\hbar}\hat{X}\hat{B}$$

Or, the same:

$$\frac{d\hat{X}}{dt} = \frac{\partial \hat{X}}{\partial t} + \frac{i}{\hbar}\left[\hat{B}, \hat{X}\right]$$

That is what we wanted to derive.

Now we shall formulate the most important postulate determining the dynamics of a system:

THE POSTULATE 4 («the Second Newton's law») The derivative of a system momentum is equal to a force acting on a system:

$$\frac{d<\vec{p}>}{dt} = <\vec{F}> \qquad (18)$$

Let's consider the corollaries of this postulate. The force operator was introduced as the operator $-\frac{\partial U}{\partial \vec{r}}$. We have:

$$\frac{d<\hat{\vec{p}}>}{dt} = <-\frac{\partial U}{\partial \vec{r}}> \qquad (19)$$

In other words,

$$\frac{d}{dt}(\Phi_\alpha, \hat{\vec{p}}\Phi_\alpha) = (\Phi_\alpha, -\frac{\partial U}{\partial \vec{r}}\Phi_\alpha)$$

Using the equation (17), we have:

$$(\Phi_\alpha, (\frac{i}{\hbar}\hat{B}\hat{\vec{p}} + \frac{\partial \hat{\vec{p}}}{\partial t} - \frac{i}{\hbar}\hat{\vec{p}}\hat{B})\Phi_\alpha) = (\Phi_\alpha, -\frac{\partial U}{\partial \vec{r}}\Phi_\alpha)$$

The operator of momentum does not change at time (that is $\frac{\partial \hat{\vec{p}}}{\partial t} = 0$):

$$(\Phi_\alpha, (\frac{i}{\hbar}\hat{B}\hat{\vec{p}} - \frac{i}{\hbar}\hat{\vec{p}}\hat{B})\Phi_\alpha) = (\Phi_\alpha, -\frac{\partial U}{\partial \vec{r}}\Phi_\alpha)$$

$\vec{\Phi}(\vec{r}, t)$ is an arbitrary function. Therefore we receive:

$$\frac{i}{\hbar}\hat{B}\hat{\vec{p}} - \frac{i}{\hbar}\hat{\vec{p}}\hat{B} = -\frac{\partial U}{\partial \vec{r}}$$

So,

$$\frac{i}{\hbar}\left[\hat{B}, \hat{\vec{p}}\right] = -\frac{\partial U}{\partial \vec{r}} \qquad (20)$$

It follows from equation (10), that $\hat{\vec{p}} = -i\hbar\frac{\partial}{\partial \vec{r}}$. Therefore we have:

$$\frac{i}{\hbar}\left[\hat{B}, -i\hbar\frac{\partial}{\partial \vec{r}}\right] = -\frac{\partial U}{\partial \vec{r}}$$

Thus,

$$\left[\hat{B}, \frac{\partial}{\partial \vec{r}}\right] = -\frac{\partial U}{\partial \vec{r}}$$

Now we shall formulate another postulate:

THE POSTULATE 5. The momentum of a system is proportional to it's speed:

$$<\hat{\vec{p}}> = m<\hat{\vec{v}}> \qquad (21)$$

Let's rewrite the last equation:
$$(\Phi_\alpha, \hat{\vec{p}}\Phi_\alpha) = m(\Phi_\alpha, \hat{\vec{v}}\Phi_\alpha)$$
It means, that
$$(\Phi_\alpha, \hat{\vec{p}}\Phi_\alpha) = (\Phi_\alpha, m\hat{\vec{v}}\Phi_\alpha)$$
As it was mentioned above, $\vec{\Phi}(\vec{r},t)$ is arbitrary. That is why
$$\hat{\vec{p}} = m\hat{\vec{v}} \qquad (22)$$

It is easy to see, that the speed is defined as $\dfrac{d<\hat{\vec{r}}>}{dt} = <\hat{\vec{v}}>$. In other words,
$$\frac{d}{dt}(\Phi_\alpha, \hat{\vec{r}}\Phi_\alpha) = (\Phi_\alpha, \hat{\vec{v}}\Phi_\alpha)$$
It follows from equation (16) that:
$$(\Phi_\alpha, (\frac{i}{\hbar}\hat{B}\hat{\vec{r}} + \frac{\partial \hat{\vec{r}}}{\partial t} - \frac{i}{\hbar}\hat{\vec{r}}\hat{B})\Phi_\alpha) = (\Phi_\alpha, \hat{\vec{v}}\Phi_\alpha)$$

The operator $\hat{\vec{r}}$ is time-independent (see its definition). Therefore $\dfrac{\partial \hat{\vec{r}}}{\partial t} = 0$. Then we receive:
$$(\Phi_\alpha, (\frac{i}{\hbar}\hat{B}\hat{\vec{r}} - \frac{i}{\hbar}\hat{\vec{r}}\hat{B})\Phi_\alpha) = (\Phi_\alpha, \hat{\vec{v}}\Phi_\alpha)$$
Using the equation (22), we have:
$$(\Phi_\alpha, (\frac{i}{\hbar}\hat{B}\hat{\vec{r}} - \frac{i}{\hbar}\hat{\vec{r}}\hat{B})\Phi_\alpha) = (\Phi_\alpha, \frac{\hat{\vec{p}}}{m}\Phi_\alpha)$$
It follows from the last formula that
$$\frac{i}{\hbar}\hat{B}\hat{\vec{r}} - \frac{i}{\hbar}\hat{\vec{r}}\hat{B} = \frac{\hat{\vec{p}}}{m}$$
Or, the same,
$$\frac{i}{\hbar}\left[\hat{B}, \hat{\vec{r}}\right] = \frac{\hat{\vec{p}}}{m} \qquad (23)$$

The last equation, as well as equation (20), is very important. Let's copy them:
$$\begin{cases} \dfrac{i}{\hbar}\left[\hat{B}, \hat{\vec{p}}\right] = -\dfrac{\partial U}{\partial \vec{r}} \\ \dfrac{i}{\hbar}\left[\hat{B}, \hat{\vec{r}}\right] = \dfrac{\hat{\vec{p}}}{m} \end{cases} \qquad (24)$$

This equation system can be interpreted, as a COMPLETE SET of EQUATIONS DETERMINING the OPERATOR $\hat{B}$ - THE OPERATOR OF THE SYSTEM CHANGE (see equation (15)). Actually these are equations which describe the evolution of coordinate and the momentum of a system. From some point of view it corresponds to the spirit of classics: the change of a system is determined the evolution of a coordinate and evolution of a momentum.

Let's solve the system (24). It is easy to see, that it is a system of linear equations with an inhomogeneous right part. The general theory says that the solution is a SUM of the common solution of a homogeneous system and a particular solution of an inhomogeneous system.

Let's find at first the particular solution of an inhomogeneous system

$$\begin{cases} \dfrac{i}{\hbar}[\hat{B},\hat{\vec{p}}] = -\dfrac{\partial U}{\partial \vec{r}} \\ \dfrac{i}{\hbar}[\hat{B},\hat{\vec{r}}] = \dfrac{\hat{\vec{p}}}{m} \end{cases}$$

The solution is a well known Hamiltonian

$$\hat{B} = -\frac{\hbar^2}{2m}\Delta + U(\vec{r}) \qquad (25)$$

Now we shall find the common solution of a homogeneous equation set:

$$\begin{cases} \dfrac{i}{\hbar}[\hat{B},\hat{\vec{p}}] = 0 \\ \dfrac{i}{\hbar}[\hat{B},\hat{\vec{r}}] = 0 \end{cases} \qquad (26)$$

Let's rewrite our system:

$$\begin{cases} [\hat{B},\hat{\vec{p}}] = 0 \\ [\hat{B},\hat{\vec{r}}] = 0 \end{cases} \qquad (27)$$

We see, that the operator $\hat{B}$ commutes with the operator of coordinate and operator of momentum. It means, that the operator $\hat{B}$ and $\hat{\vec{p}}$ have a complete set of the common eigenvectors. Also for there is a complete set from the common eigenvectors for $\hat{B}$ and $\hat{\vec{r}}$.

Let's solve this system in a one dimensional case. Our solution will be very easy to generalize in a multidimensional case. It is clear, that the full eigenvector set for the coordinate operator is a as following: $\Psi_a = \delta(x-a)$. We have mentioned that these eigenvectors must be the eigenvectors for $\hat{B}$. That is,

$$\hat{B}\delta(x-a) = b(a)\delta(x-a) \qquad (28)$$

As known, any function $f(x)$ can be decomposed on $\delta$-functions:

$$f(x) = \int f(a)\delta(x-a)\,da$$

Therefore to calculate the action of operator $\hat{B}$ over any to functions, it is enough to decompose it in $\delta$-functions:

$$\hat{B}f(x) = \hat{B}\int f(a)\delta(x-a)\,da$$

Or,

$$\hat{B}f(x) = \int \hat{B}[f(a)\delta(x-a)]\,da$$

Let's use the linearity:

$$\hat{B}f(x) = \int f(a)\hat{B}\delta(x-a)\,da$$

From (28) we can write as following:

$$\hat{B}f(x) = \int f(a)b(a)\delta(x-a)\,da$$

Then according to the properties of $\delta$-function we see:

$$\hat{B}f(x) = b(x)f(x) \qquad (29)$$

The last equation is important. IT STATES, THAT THE OPERATOR $\hat{B}$ IS AN OPERATOR, MULTIPLYING ON ARBITRARY FUNCTION $f(x)$.

As we have understood, the eigenvectors of operator $\hat{B}$ must be the eigenvectors of the momentum operator. That is,

$$\hat{B} e^{ikx} = \alpha(k) e^{ikx} \tag{30}$$

Using equation (29), we receive:

$b(x) e^{ikx} = \alpha(k) e^{ikx}$

Or,

$b(x) = \alpha(k)$

It means, that $b(x)$ and $\alpha(k)$ are constants:

$b(x) = \alpha(k) = C$

Substituting this equation in (29), we receive:

$$\hat{B} f(x) = C f(x) \tag{31}$$

THE LAST FORMULA SAYS US THAT THE OPERATOR $\hat{B}$ IS AN OPERATOR OF MULTIPLYING BY A CONSTANT.

These reasoning are extended in a multidimensional case. Now let us summarize a little bit, what we have done. We have written a set of equations (24) for the operator $\hat{B}$ - operator of the system change. We have understood, that it's the common solution is a sum of a particular solution of inhomogeneous system and common solution of a homogeneous system.

We have found particular solution in (25). It is:

$$\hat{B}_1 = -\frac{\hbar^2}{2m}\Delta + U(\vec{r})$$

The common solution of a homogeneous system is (see (31)):

$$\hat{B}_2 = C$$

Thereby operator $\hat{B}$ is the sum:

$$\hat{B} = -\frac{\hbar^2}{2m}\Delta + U(\vec{r}) + C \tag{32}$$

The constant can be included in a potential energy:

$$\hat{B} = -\frac{\hbar^2}{2m}\Delta + U(\vec{r}) \tag{33}$$

Let's rewrite this equation, using (15):

$$i\hbar \frac{\partial \vec{\Phi}}{\partial t} = \hat{B}\vec{\Phi}, \quad \text{where} \quad \hat{B} = -\frac{\hbar^2}{2m}\Delta + U(\vec{r}) \tag{34}$$

It also is a SCHRODINGER EQUATION, which we wanted to prove.

We have achieved our aim. Let's remark, that $\vec{\Phi}(\vec{r},t)$ is a vector function. In our theory each component of this function changes the same way. In our proof we did not consider the magnetic field. The spin may be taken into account. Let's remark, that we at all did not fix the number of components of this vector function. We have obtained the main equation, but there is very much to do.

### The Appendix 1

There is a statement:

**If $e^{\hat{A}t}$ is a unitary operator, then the operator $\hat{A}$ is antihermittean (that is $\hat{A}^* = -\hat{A}$).**

**The proof:** If $e^{\hat{A}t}$ is unitary, then $e^{\hat{A}t}(e^{\hat{A}t})^* = 1$.

But $(e^{\hat{A}t})^* = e^{\hat{A}^* t}$. Therefore we have:
$$e^{\hat{A}t} e^{\hat{A}^* t} = 1$$
Let us take a time derivative:
$$\hat{A} e^{\hat{A}t} e^{\hat{A}^* t} + e^{\hat{A}t} \hat{A}^* e^{\hat{A}^* t} = 0$$
Now shall we multiply both parts on $e^{-\hat{A}^* t}$:
$$\hat{A} e^{\hat{A}t} + e^{\hat{A}t} \hat{A}^* = 0$$
The operators $\hat{A}t$ and $e^{\hat{A}t}$ commute:
$$e^{\hat{A}t} \hat{A} + e^{\hat{A}t} \hat{A}^* = 0$$
Now let's multiply both parts on $e^{-\hat{A}t}$:
$$\hat{A} + \hat{A}^* = 0$$
Or, the same,
$$\hat{A}^* = -\hat{A}$$
That is what we wanted to prove.

## The appendix 2.

When we received the Schrödinger equation, we have made one proposal. We stated that the change operator $\hat{A} = \frac{1}{i\hbar}\hat{H}$ is time-independent - see equalities (13), (14), (15). Let's remove the first limitation and consider, that the operator $\hat{A}$ changes in time:
$$\hat{A} = \hat{A}(t) \tag{1}$$
Then the proof will change. Particularly, we can not write, that
$$\vec{\Phi} = e^{\hat{A}t}\vec{\Phi}_0$$
And it is impossible to prove this way that the operator $\hat{A}$ is anti-hermittean and that the system change operator $\hat{B}$ is hermitean. Therefore we shall go other way. Let us introduce the evolution operator on the basis of Postulate 1. We know that the system is completely determined by a function $\vec{\Phi}$. Therefore, knowing $\vec{\Phi}$ at any arbitrary moment of time $t_1$, we know it in any other moment $t_2$. Therefore the operator which establishes up a correspondence between the function $\vec{\Phi}(\vec{r},t_1)$ in and $\vec{\Phi}(\vec{r},t_2)$ is called an evolution operator (we must never mix up the evolution operator and the system change operator $\hat{B}$):
$$\vec{\Phi}(\vec{r},t_2) = U(t_1,t_2)\vec{\Phi}(\vec{r},t_1) \tag{2}$$
This operator has a plenty of apparent properties:
1. The evolution operator is unitary (because it saves the norm - the argumentation is the same).
2. The evolution operator is inversible. Indeed, we see that operator, inversed to $U(t_1,t_2)$ is the operator $U(t_2,t_1)$.
3. The evolution operator satisfies the property

$$U(t_1,t_2)U(t_2,t_3) = U(t_1,t_3)$$

Indeed the system evolution from the moment $t_1$ to the moment $t_3$ can be represented, as the evolution from the moment $t_1$ to the moment $t_2$, and then from the moment $t_2$ to the moment $t_3$.

Now it is easy to calculate, that the function $\vec{\Phi}$ is obeys an equation

$$i\hbar\frac{\partial \vec{\Phi}}{\partial t_2} = \hat{B}\vec{\Phi}, \text{ where } \hat{B} = i\hbar\frac{\partial U(t_1,t_2)}{\partial t_2}U^{-1}$$

And, thereby we easily come to a standard equation of the system change the first order type. (see (12) – there we obtained it in a rather more complicated and unstrict way). In addition we have the unicity of the solution, if initial conditions are given - that is a well known first order equation property.

Now we can easily prove the self-conjugacy of the operator $\hat{B}$. If we substitute the equation (2) in the last expression, we receive:

$$i\hbar\frac{\partial}{\partial t_2}U(t_1,t_2)\vec{\Phi}(\vec{r},t_1) = \hat{B}U(t_1,t_2)\vec{\Phi}(\vec{r},t_1) \quad (3)$$

The function $\vec{\Phi}(\vec{r},t_1)$ is arbitrary. So we have:

$$i\hbar\frac{\partial}{\partial t_2}U(t_1,t_2) = \hat{B}U(t_1,t_2) \quad (4)$$

The operator $\hat{B}$ is inversible. That is why

$$i\hbar\frac{\partial U}{\partial t_2}U^{-1} = \hat{B} \quad (5)$$

Then

$$\hat{B}^* = -i\hbar(U^{-1})^*\left(\frac{\partial U}{\partial t_2}\right)^*$$

It follows from the unitarity of the evolution operator that $(U^{-1})^* = U$ (in fact, the unitarity means, that $UU^* = 1$, or, that same, $(U^{-1})^* = U$). Therefore we easily receive:

$$\hat{B}^* = -i\hbar U\left(\frac{\partial U}{\partial t_2}\right)^*$$

Or,

$$\hat{B}^* = -i\hbar U\frac{\partial U^*}{\partial t_2}$$

The unitarity also gives us the condition $U^* = U^{-1}$. Then

$$\hat{B}^* = -i\hbar U\frac{\partial U^{-1}}{\partial t_2} \quad (6)$$

Now we shall compare the equations (5) and (6):

$$\hat{B} - \hat{B}^* = i\hbar\frac{\partial U}{\partial t_2}U^{-1} - i\hbar U\frac{\partial U^{-1}}{\partial t_2} = i\hbar(\frac{\partial U}{\partial t_2}U^{-1} - U\frac{\partial U^{-1}}{\partial t_2})$$

It follows from the formula $\frac{\partial}{\partial t}(ab) = \frac{\partial a}{\partial t}b + a\frac{\partial b}{\partial t}$ that

$$\hat{B} - \hat{B}^* = i\hbar \frac{\partial}{\partial t_2}(UU^{-1})$$

Therefore

$$\hat{B} - \hat{B}^* = i\hbar \frac{\partial}{\partial t_2}(1)$$

In other words,

$$\hat{B} = \hat{B}^*$$

We see that operator $\hat{B}$ is self-conjugated. That is the end of the proof. We come to an obvious fact: self-conjugacy of a Hamiltonian does not depend on the external influence. It may be constant or it may vary in time.

That was exactly the goal of the article. We have derived the Schrödinger equation but as it was possible to prove, that it remains same if the parameters change in time(for example, value of an electrical field and so on).

### **The literature.**